\documentstyle[aps,prb,graphicx,floats]{revtex}

\input epsf         
\epsfverbosetrue

\begin{document}
\twocolumn[\hsize\textwidth\columnwidth\hsize\csname@twocolumnfalse%
\endcsname

\draft

\title{Weak Pseudogap in  Crystals of Pb$_{2}$ Sr$_{2}$(Y,Ca)Cu$_{3}$O$_{8+\delta}$}

\author{Ya-Wei Hsueh and B. W. Statt}
\address{Department of Physics, University of Toronto,
         Toronto, Ontario, Canada M5S 1A7} 
\author{M. Reedyk}
\address{Department of Physics, Brock University,
         St. Catharines, Ontario, Canada L2S 3A1}
\author{J. S. Xue and J. E. Greedan}
\address{Brockhouse Institute for Material Research, McMaster University,
         Hamilton, Ontario,}
\address{Canada L8S 4M1}

\date{\today}

\maketitle

\begin{abstract} 
We report on NMR  measurements in  underdoped Pb$_{2}$ Sr$_{2}$(Y,Ca)Cu$_{3}$O$_{8+\delta}$
crystals. A pseudogap is observed in the Knight shift and spin--lattice relaxation rate. In contrast
to other underdoped compounds, the pseudogap observed in the Knight shift is weak and occurs
at a significantly lower temperature. On the other hand, the effect the pseudogap has on
spin--lattice relaxation  is quite similar to that in other compounds. The contrast between
weak and strong pseudogaps is discussed.
\end{abstract}

\vskip 0.1 true in

]



An unusual feature of the normal-state properties of some high--$T_c$
superconductors is the  pseudogap. This manifests itself,
for example, in the
spin susceptibility which decreases with decreasing temperature.
The pseudogap has been observed in underdoped systems such as
YBa$_{2}$Cu$_{3}$O$_{6.63}$\cite{Taki91-1,AOM89,Wal90} (YBCO6.6) and
YBa$_{2}$Cu$_{4}$O$_{8}$\cite{BMRB94} (Y248)
but not in optimally-doped compounds, such as YBa$_{2}$Cu$_{3}$O$_{7}$\cite{Taki90} (YBCO7),
which have temperature independent susceptibilities.

A pseudogap appears not only in spin properties, but also in charge transport.
The c-axis optical conductivities of several underdoped materials, YBCO6.6
and Y248,  show a gap-like feature in
the normal state\cite{Timusk95-1} below about 300-400 cm$^{-1}$.
Puchkov {\it et al.}\cite{Puchkov} have systematically analyzed the evolution of the scattering
rate 1/$\tau$($\omega$,$T$) in infrared reflectivity in a series of compounds from
underdoped to overdoped.
In the underdoped region they see a gap-like depression (pseudogap) in 1/$\tau$($\omega$,$T$)
opening up below $T \simeq$140--160 K, while in the overdoped region they do not observe a gap.
DC resistivity measurements also display signs of a pseudogap.
The resistivity for YBCO7 follows a linear temperature dependence
over a very wide temperature range above $T_c$. 
On the other hand the resistivity of Y248 deviates from
linearity at a temperature well above $T_c$\cite{Bucher}, consistent with the pseudogap observed
in other measurements.

Nuclear magnetic resonance (NMR) has been used extensively as a microscopic
probe of the electronic properties of high-$T_c$ compounds.
The Knight shift $K_s(T)$ is proportional to the spin susceptibility
$\chi^{\prime}({\bf q}=0, \omega\simeq0)$.
The spin--lattice relaxation rate $T_1^{-1}$ averages over all {\bf q} of
$\chi^{\prime\prime}({\bf q}, \omega)$.
As $\chi^{\prime\prime}({\bf q}, \omega)$ is strongly peaked at the antiferromagnetic
wavevector {\bf Q}=($\pi$/a,$\pi$/a), $T_1^{-1}$ of the planar Cu nuclei
predominantly probes $\chi^{\prime\prime}({\bf Q})$.
Although both of these quantities develop a pseudogap they do so at
quite different temperatures. As we will discuss later, the pseudogap manifests
itself differently at {\bf q}=0 and {\bf q}={\bf Q}.

To date there is no consensus on the origin of the pseudogap.
Emery and Kivelson\cite{EK95} suggest that pair formation and phase coherence do not 
necessarily occur at the same temperature.
The lower carrier density in the underdoped region may result in pair formation without
condensation into a coherent state at temperatures well above $T_c$, thereby
producing a pseudogap.
Antiferromagnetic (AF) spin fluctuations and the evolution of the Fermi surface on doping
may also contribute to the creation of the pseudogap\cite{Chub}.
More exotic theories involving charge--spin separation have also been considered, with interplanar
coupling, as a source of the pseudogap\cite{Alt}.
Recently Atkinson, Wu and Carbotte\cite{Atkinson} have proposed that the pseudogap is not
actually a spin gap but a result of strong AF spin fluctuations and interband coupling.

There have been few reports of magnetic resonance studies on
 Pb$_{2}$Sr$_{2}$Y$_{1-x}$Ca$_x$Cu$_{3}$O$_{8+\delta}$ (PSYCCO).
So far studies have been restricted to optimally doped powder samples where $x$ = 0.5.
Spin-lattice relaxation measurements\cite{Koh92,Yos91} of Cu spins in the CuO$_{2}$ planes
are similar to those observed in   YBCO7.
The Pb NMR Knight shift\cite{Koh92} in PSYCCO
is temperature independent in the normal state.
Both of these temperature dependences are consistent with optimally doped samples with
no hint of a pseudogap.

In this paper, we present measurements of the $^{63}$Cu Knight shifts for the planar  Cu(2)
sites in both the superconducting and normal state of slightly underdoped
PSYCCO crystals ($x \lesssim 0.5$).
The nuclear spin--lattice relaxation rate $T_1^{-1}$ was also measured in the
normal state. 
We find evidence of a weak pseudogap developing in the Knight shift at considerably
lower temperatures  than usual. In contrast, the pseudogap appearing in $T_1^{-1}$
occurs much closer in  temperature to that of the strong pseudogap compounds.


Preparation of the single crystals of 
PSYCCO used here has been described by Xue {\it et al.}\cite{Xue90}.
Magnetic dc susceptibility measurements reveal the onset of $T_c$ to be 
80 K in a field of 5 Oe.
Our sample consists of about 300 c--axes aligned single crystals, each of dimensions $\sim$
0.5mm$\times$ 0.5mm$\times$ 0.3mm.
These crystals are slightly underdoped as indicated in
the resistivity\cite{Reedyk}, showing a change in slope around 150 K. As expected for
underdoped crystals, the magnitude of the resistivity is greater than that of the optimally
doped samples\cite{Xue90} with a $T_c$ of 84 K.
Measurements of the optical conductivity also indicate the presence of a pseudogap\cite{Reedyk}.
Our NMR experiments were carried out with a standard pulsed spectrometer.

\begin{figure}
        \begin{center}
                \leavevmode
                \includegraphics[origin=c, angle=-90, width=8cm, clip]{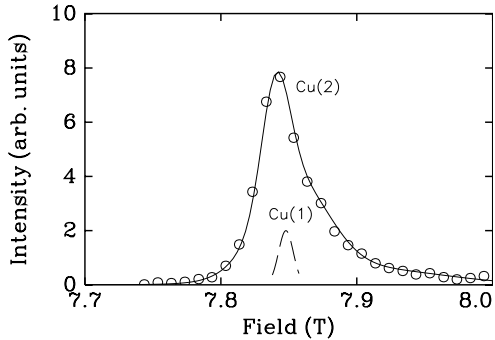}
        \end{center}

\caption{$^{63}$Cu(2) central transition taken at 90.6 MHz with {\bf c}$\perp${\bf H$_0$}.
Dashed line indicates the position of the $^{63}$Cu(1) line which is saturated in this spectrum.}
\label{fig1}
\bigskip
\end{figure}

The NMR lineshape consists of two overlapping lines, one from the planar Cu(2) sites
and the other from the Cu(1) sites between the PbO layers. As the spin--relaxation rate
for Cu(1) is four orders of magnitude smaller\cite{YaWei2} than that for Cu(2) we
conducted our experiments at a repetition rate sufficient to saturate the Cu(1) signal.
Figure\ \ref{fig1} displays the Cu(2) central transition with {\bf c}$\perp${\bf H$_0$}. The
asymmetric lineshape is due to the angular spread amongst the crystal mosaic c--axes.
The large linewidth is due to the distribution of electric field gradients
from the Ca$^{+2}$/Y$^{+3}$ layer between the Cu(2)O$_2$ bilayer.
Also shown in Fig.\ \ref{fig1} is the Cu(1) central transition, taken with a much lower
repetition rate.

The Knight shift was obtained from the peak line position by subtracting the
second--order quadrupolar shift. To second order the resonance frequency $\nu_0$
for {\bf c}$\perp${\bf H$_0$}  is given by
\FL
\begin{equation}
        \nu_0 = \nu_L + (3/16)\nu_Q^2/\nu_L   \label{eq1}
\end{equation}
where $\nu_L$ is the Larmor frequency and $\nu_Q$ is the quadrupolar frequency.
Our NQR spectrum, used to determine $\nu_Q$, is similar to those obtained
previously\cite{Koh92,Yos91}.
This procedure of subtracting the second--order shift was checked at two temperatures
by measuring the field dependence of the resonance frequency. One can independently
extract the quadrupolar shift from the Knight shift by plotting the effective Knight
shift $K_{\text{eff}}$ versus $[(1 + K_{\text{eff}})/\nu_L]^2$. The resulting intercept is equal to
the Knight shift $K$.

\begin{figure}
        \begin{center}
                \leavevmode
                \includegraphics[width=8cm, clip]{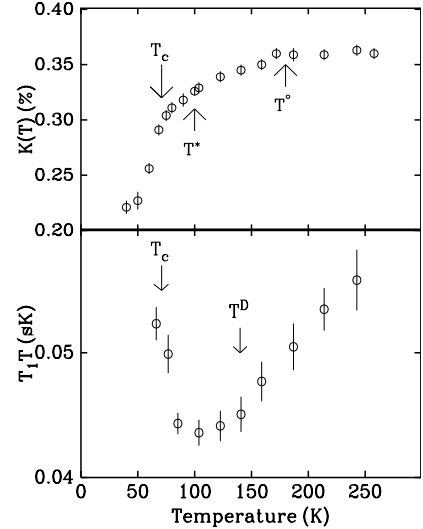}
        \end{center}
\caption{(a) $^{63}$Cu(2) Knight shift for PSYCCO with {\bf c}$\perp${\bf H$_0$}.
The arrow indicates the value of $T_c$ at H$_0$=8T. 
 (b) Spin--lattice relaxation time for PSYCCO. $T^{\circ}$, $T^{\star}$ and $T^D$
are described in the text.}
\label{fig2}
\bigskip
\end{figure}

The Knight shift consists of two parts, the temperature independent orbital
component $K_{\text{orb}}$ and the spin component  $K_s(T)$.
The spin Knight shift is proportional to the spin susceptibility $\chi_s(T)$
and the hyperfine coupling constants.
In Figure\ \ref{fig2}a we show the temperature dependence of the Knight shift for PSYCCO.
$K_s$ is temperature independent at high temperatures but slowly decreases linearly
with temperature at $T^{\circ} \simeq$ 180 K. Starting at $T^{\star} \simeq$ 100 K the
Knight shift gradually decreases faster until at $T_c$ it enters the superconducting
state.
Below $T_c$,  $K_s$ decreases more rapidly, now with a positive curvature.
Note that at $T_c$, $K_s$ has dropped to  one half of the total excursion in $K_s$
(0.14\%) from high temperature to $T=0$.

\begin{figure}
        \begin{center}
                \leavevmode
                \includegraphics[width=8cm, clip]{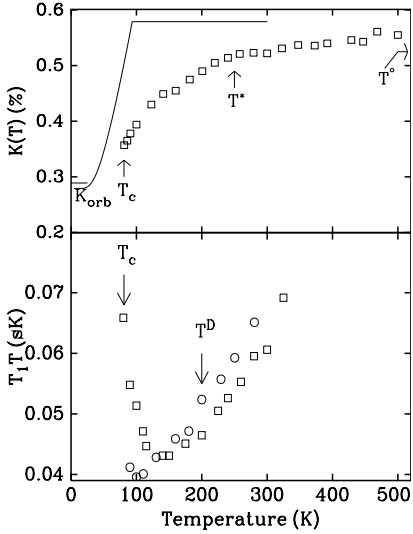}
        \end{center}
\caption{(a) $^{63}$Cu(2) Knight shift for Y248 (squares)\cite{BMRB94} with {\bf c}$\perp${\bf
H$_{0}$}. The solid line represents data for optimally doped YBCO7.
 (b) Spin--lattice relaxation time for Y248 (squares) and YBCO7 (circles)\cite{Taki91-1}.}
\label{fig3}
\bigskip
\end{figure}

The Knight shift suggests that a weak pseudogap starts to develop at $T^{\circ}$.
This is to be compared to the strong pseudogap of
Y248\cite{BMRB94} which opens up at $T^{\circ} \geq$ 500 K
as shown in Fig.\ \ref{fig3}a.
Here the total excursion in $K_s$ is 0.26\% with 80\% of that occurring in the normal
state. Thus the drop in $K_s$ in the superconducting state is quite small compared to
that of an optimally doped sample (see Fig.\ \ref{fig3}a).
A similarly strong pseudogap  is observed in
YBCO6.6\cite{Taki91-1,AOM89,Wal90}.
In this sense one might say that the  pseudogap observed in PSYCCO is not
as fully developed as the strong pseudogap seen in the two yttrium compounds.


Presented in Fig.\ \ref{fig2}b is $T_1T$ for {\bf c}$\perp${\bf H$_{0}$}.
Typical of high--$T_c$ compounds $T_1T$ is not temperature independent
as one expects of a degenerate Fermi gas. At high temperatures the spin--lattice
relaxation rate is increasing with decreasing temperature due to  increasing AF spin
fluctuations. In optimally--doped compounds this continues to just above $T_c$,
as illustrated for YBCO7 in Fig.\ \ref{fig3}b. Under--doped compounds,
on the other hand, start to deviate from the linear temperature dependence
well above $T_c$. In PSYCCO this occurs at $T^D \simeq$ 140 K whereas in
Y248 $T^D \simeq$ 200 K as shown in Fig.\ \ref{fig2}b and \ref{fig3}b respectively.

The most striking feature of this work is that the Knight shift of weak and strong
pseudogap materials are quite different while their spin--lattice relaxation rates are
quite similar. Thus the manifestation of the pseudogap at {\bf q}=0 is quite different
than at {\bf q}={\bf Q}. In order to elucidate this situation let us consider the resistivity
measurements on PSYCCO and Y248. In both cases the linear $T$ resistivity observed at
high temperatures changes at $T \simeq$ 160 K to a more rapidly decreasing resistivity
with temperature. This is quite similar to the behaviour of $1/T_1T$ at $T^D$. Both
the spin--lattice relaxation and resistivity are dominated by AF spin fluctuations\cite{StattG}.
Thus it is not surprising that the effect the pseudogap has on the AF spin fluctuation spectrum
manifests itself at a similar temperature for both spin--lattice relaxation and resistivity.

The Knight shift though is a measure of the long wavelength spin susceptibility
which may be affected by the pseudogap in a different manner. In the language of
the MMP model $\chi^{\prime}(0)$ describes the quasiparticles\cite{Chub}. In contrast to
$1/T_1T$ and resistivity measurements, the onset temperatures for the weak and
strong pseudogap materials are quite different. In both cases though, $T^{\circ}$ is
greater than $T^D$. But in PSYCCO $T^{\star}$ is less than $T^D$ and in Y248
$T^{\star}$ is greater than $T^D$ .

Another interesting correlation between the weak and strong pseudogap materials
is that the weak pseudogap develops at a much lower temperature {\bf and} the change in
magnitude of
the Knight shift down to $T_c$ is much less than that for the strong pseudogap.
Thus it seems that the weak pseudogap is not as developed as the strong pseudogap
at $T_c$. Perhaps this is due to the fact that $T^{\circ}$ is much closer to $T_c$ in the
weak pseudogap material and that had $T_c$ been proportionally reduced, the Knight
shift would have been more heavily suppressed at $T_c$. Among the various underdoped
compounds, PSYCCO is unique in this respect.

This behaviour is echoed in the c--axis optical conductivity\cite{Timusk95-1}. The low frequency
conductivity for YBCO6.6 at $T_c$ is less than 20\% of the high temperature limit.
This is consistent with the behaviour of the Knight shift (similar to that in Fig.\ \ref{fig3}a).
On the other hand the optical conductivity for PSYCCO has only decreased by 50\%
at $T_c$\cite{Reedyk}, again similar to the drop displayed in the Knight shift.
Thus the c--axis optical conductivity and Knight shift both illustrate the weak nature
of the pseudogap in PSYCCO.

One would expect $T^{\circ}$ and $T^{\star}$ to be correlated with the energy at which the pseudogap
opens up in the optical spectrum if the pseudogap was indeed due to a gap in the
excitation spectrum. Basov {\it et al.}\cite{Basov} have found the pseudogap to develop
at about 300--400 cm$^{-1}$ in YBCO6.6 and Y248. These compounds both have $T^{\circ} \geq$
500 K. On the other hand PSYCCO has a much lower $T^{\circ}$ of 180 K with the
pseudogap still well developed\cite{Reedyk} at 700 cm$^{-1}$. Thus it seems that the
pseudogap's onset temperature $T^{\circ}$ is either negatively correlated to the onset
frequency or perhaps not at all.

Reedyk {\it et al.}\cite{Reedyk} have noted that there is no correlation between the
magnitude of the high frequency c--axis conductivity and the pseudogap amongst the
various underdoped materials. There also appears to be no correlation with the
carrier concentration. The plasma frequency $\omega_p$ for YBCO6.6 is 0.8 eV which
is similar to the value for PSYCCO of 0.6 eV. These underdoped compounds both have
$\omega_p$ considerably less than the value for optimally doped YBCO where
$\omega_p$ = 1.4 eV.


In summary, we have observed a weak pseudogap in underdoped PSYCCO crystals.
The effect the pseudogap has on the Knight shift in strongly dependent on whether
the pseudogap is strong or weak. In contrast, the effect the pseudogap has on the AF
spin fluctuation spectrum, as measured by spin--lattice relaxation and resistivity,
is almost independent of the pseudogap strength.

\acknowledgements
We thank Thom Mason for the use of his 18T magnet for preliminary measurements.
This work was supported by a research grant from the Natural Sciences and Engineering
Research Council of Canada.







\end{document}